\documentclass{article}
\usepackage{spconf,amsmath,graphicx}
\usepackage{amsfonts,amssymb}
\usepackage{multirow}
\usepackage{graphicx}
\usepackage{caption}
\usepackage{flushend}
\usepackage{bm}
\usepackage{booktabs}
\usepackage{subfigure}

\usepackage{color, colortbl}
\definecolor{Gray}{gray}{0.93}

\DeclareMathOperator{\xb}{\mathbf{x}}
\DeclareMathOperator{\ab}{\mathbf{a}}
\DeclareMathOperator{\tb}{\mathbf{t}}

\newcommand{\Hmat}{{\bf H}}

\newcommand{\sv}{{\boldsymbol{s}}}
\newcommand{\zv}{{\boldsymbol{z}}}
\newcommand{\hv}{{\boldsymbol{h}}}
\newcommand{\thetav}{{\boldsymbol{\theta}}}

\newcommand{\RN}[1]{%
	\textup{\lowercase\expandafter{\it \romannumeral#1}}%
}

\newcommand{\tts}[1]{{\small\texttt{#1}}}

\title{Music Era Recognition Using Supervised Contrastive Learning and Artist Information}
\name{Qiqi He$^{2*}$\thanks{$^*$The author conducted this work as an intern at ByteDance.}, Xuchen Song$^1$, Weituo Hao$^1$, Ju-Chiang Wang$^1$, Wei-Tsung Lu$^1$, Wei Li$^2$}
\address{$^1$ ByteDance\\$^2$ School of Computer Science and Technology, Fudan University, Shanghai, China\\
\tt\small \{heqq20, weili-fudan\}@fudan.edu.cn\\
\tt\small \{xuchen.song, weituohao, ju-chiang.wang, weitsung.lu\}@bytedance.com
}

\begin{document}
\ninept

\maketitle

\begin{abstract}
Does popular music from the 60s sound different than that of the 90s? 
Prior study has shown that there would exist some variations of patterns and regularities related to instrumentation changes and growing loudness across multi-decadal trends. This indicates that perceiving the era of a song from musical features such as audio and artist information is possible. Music era information can be an important feature for playlist generation and recommendation. However, the release year of a song can be inaccessible in many circumstances. 
This paper addresses a novel task of music era recognition.
We formulate the task as a music classification problem and propose solutions based on supervised contrastive learning. An audio-based model is developed to predict the era from audio. For the case where the artist information is available, we extend the audio-based model to take multimodal inputs and develop a framework, called \textbf{M}ulti\textbf{M}odal \textbf{C}ontrastive (MMC) learning, to enhance the training. Experimental result on Million Song Dataset demonstrates that the audio-based model achieves 54\% in accuracy with a tolerance of 3-years range; incorporating the artist information with the MMC framework for training leads to 9\% improvement further. 

\end{abstract}

\section{Introduction}
\label{sec:intro}

As a central goal of the popular music industry over the past 60 years, music is created to be attractive to listeners 
\cite{ball2010music,huron2008sweet}. It is believed that music must incorporate variations of patterns and regularities that build upon people's expectations and memories across years \cite{huron2008sweet,honing2017musical}.	
Although scientific evidence still remains less specific, prior study has shown that these variations could be related to instrumentation changes and growing loudness levels \cite{serra2012measuring}. Timbral and mood variations were also found across multi-decadal trends in \cite{interiano2018musical}.
Inspired by these prior findings, developing a predictive model to recognize the music era from audio can be a potentially feasible task.
	
	
In music categorization, people typically break down music era by decades (e.g., 80s, 90s, and 2000s), which are commonly used as tags in major music streaming services such as Spotify and Pandora. A song's release year can carry a meaningful context of culture, mood, a time in life, or a peek into history, offering a straightforward way to organize songs for use such as playlist generation and recommendation \cite{pohle2007reinventing, schedl2012model}. 
Although the music era can be inferred via the release year, estimating the era for a song from audio is practically useful in various scenarios.
As the Internet has become ubiquitous, users' content sharing and derivative work have grown rapidly on media platforms such as TikTok and YouTube. These activities may involve music reuse or edits (e.g., excerpt and remix) which can cause loss of the original metadata. Moreover, the cover version of an old tune recorded recently would also yield confusion.


	
	
In this paper, we introduce the music era recognition task, which can be formulated as an application-specific music classification problem that aims to classify songs into different \emph{year ranges} (e.g., decades), depending on the granularity. A major challenge is to differentiate songs from nearby year ranges, because the variation between two songs can be less notable if they were released within a range of years. In addition, we also observe an imbalanced distribution of songs across years in data. Therefore, our design principle requires the model to learn representations near each other for two song inputs if both are in the same year range, and otherwise far apart. 
For this purpose, we adopt the supervised contrastive learning framework \cite{khosla2020supervised}, which has shown state-of-the-art performance in image classification. To incorporate artist information, we introduce the MultiModal Contrastive (MMC) learning framework, which includes a novel architecture for multimodal inputs and an unsupervised contrastive loss for learning a robust combination of the audio and artist embeddings. The MMC loss can force clustering the embeddings of the songs from the same artist \cite{chen2020simple}.


To our knowledge, this work represents the first attempt to develop automatic methods for music era recognition. 
From a technical perspective, our work is related to music auto-tagging. Early approaches to music tagging include methods using the context information of music to predict the user preference tags \cite{10.1145/957013.957040}. Recent advances in deep learning accelerated the development of content-based music tagging technology \cite{nam2018deep}. Numerous systems based on convolutional neural networks (CNN) were proposed \cite{choi2016automatic, pons2017end, won2020data}. However, little attention was paid to music era recognition, possibly because the song release year is typically accessible through metadata in common use cases such as streaming. Nevertheless, as aforementioned, era information can be missing in many circumstances.   

	
As the sounds of an era could be better defined by the popular songs of the era, we consider this work is also related to the task of \emph{hit song prediction} \cite{zangerle2019hit}, where the goal is to predict if a song would be successful/popular within a period of (future) time based on its musical features. Two types of features were explored: \emph{internal} features extracted from audio \cite{yang2017revisiting,yu2017hit} and \emph{external} ecosystem-related features such as social media and market data \cite{interiano2018musical}.
It is also found that adding an artist factor (i.e., if the artist had succeeded in the near past) can significantly improve the prediction accuracy \cite{interiano2018musical}.


To summarize our technical contributions, we propose three variants of methods, including:
(1) \emph{Audio-CNN}: a CNN model that predicts music era from audio (Section 2.1); 
(2) \emph{Audio-SUC}: an enhanced model that incorporates supervised contrastive learning to improve era recognition from audio (Section 2.2);
(3) \emph{AudioArt-MMC}: an augmented model of \emph{Audio-SUC} that joints the audio and artist embeddings, and is trained with an additional MMC loss (Section 2.4). We evaluate the proposed methods on two datasets (Section 3). Our results show that \emph{Audio-SUC} significantly outperforms \emph{Audio-CNN}, 
and incorporating the artist's information with the MMC loss (\emph{AudioArt-MMC}) further improves the performance.


	
    
    
    
    

    
    
    
    
    
    

\section{Proposed Method}

\subsection{Convolutional Neural Network (CNN)}
\label{sec:cnn}

Due to insufficient research on music era recognition, we develop \emph{Audio-CNN}, a modified CNN proposed in \cite{ibrahim2020should} as our baseline model, which contains a stack of CNN layers as a filter to extract local features, followed by a linear layer as to output classification logits. The feature map of $l$-th CNN layer is computed as follows:
\begin{align} \label{eq:conv_layer} 
   \Hmat_{\ell} = \text{ELU}(\text{BN}(\text{CNN}(\Hmat_{\ell -1}))),
\end{align}
where BN is the batch normalization operation, and ELU is the exponential linear unit. Each convolution is operated based on a kernel size of $3 \times 3$, followed by an average pooling layer. The model is trained with cross-entropy loss: 
\begin{align}
\mathcal L_{\mathrm{MLE}}= - \frac{1}{N} \sum_{i=1}^N y_{i} \cdot \log
(f(\xb_{i})), 
\end{align}
where $N$ is the number of songs, $\xb_i$ is the mel-spectrogram extracted from the audio of song $i$, and $y_{i}$ is the corresponding one-hot encoded label out of $C$ possible era classes.

    
\begin{figure}[t]
  \vspace{-2mm}
  \subfigure[\label{fig:bef_sup}without EC loss.]{%
      \includegraphics[width=0.24\textwidth]{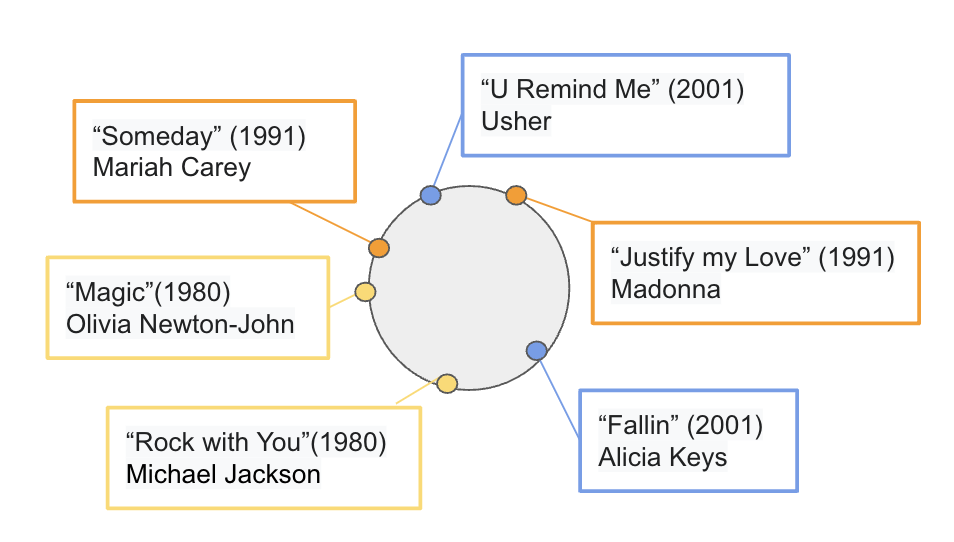}}
\hspace{\fill}
  \subfigure[\label{fig:aft_sup}with EC loss.]{%
      \includegraphics[width=0.24\textwidth]{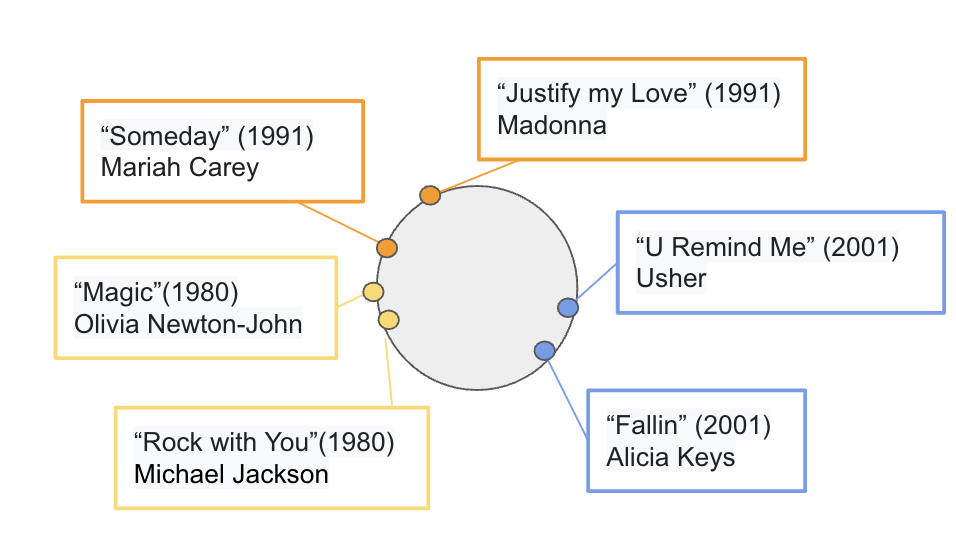}}    
\hspace{\fill}
\vspace{-2mm} \\

  \subfigure[\label{fig:bef_un}without MMC loss.]{%
      \includegraphics[width=0.24\textwidth]{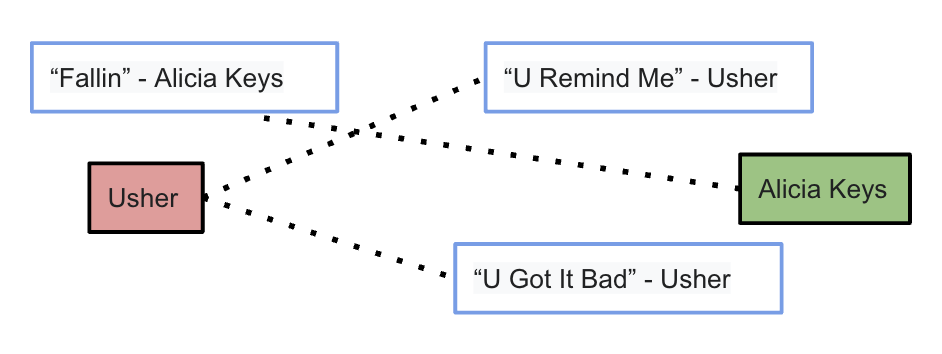}}
\hspace{\fill}
  \subfigure[\label{fig:aft_un}with MMC loss.]{%
      \includegraphics[width=0.24\textwidth]{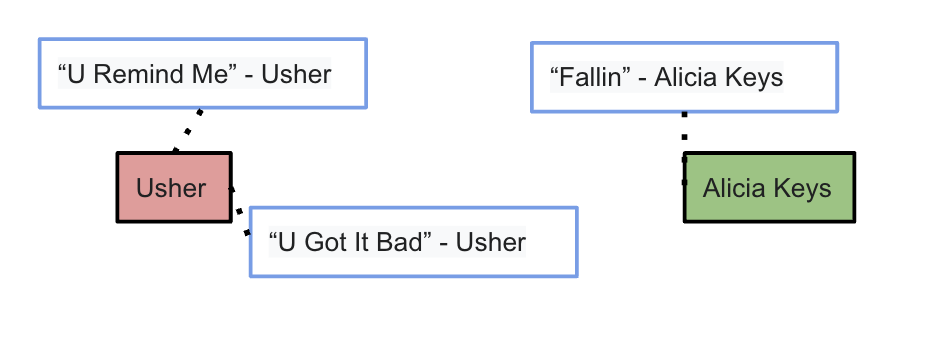}}    
\hspace{\fill}
\vspace{-2mm}

\caption{\label{fig:mmc} Illustrating the effects of contrastive learning. From (a) to (b), EC loss help aggregating songs based on the era classes. From (c) to (d), MMC loss improves clustering songs based on artists.}
\vspace{-2mm}
\end{figure}
    
\subsection{Supervised Contrastive (SUC) Learning}

Despite the large scale CNNs can be a powerful model, we consider its architecture and training scheme using cross-entropy loss are still insufficient to tackle the discrimination of nearby eras.
Therefore, we develop a novel architecture, called \emph{Audio-SUC}, to incorporate the contrastive learning \cite{khosla2020supervised,yu2022graph,marrakchi2021fighting, jiang2021self,wang2021contrastive}, which aims to learn robust representations by discriminating positive and negative embedding pairs based on the era labels. 


Specifically, we follow \cite{khosla2020supervised} to design the \emph{era contrastive} (EC) loss, which forces clusters of audio embedding belonging to the same era class to be pulled together in the embedding space, and meanwhile pushes apart the audio embeddings from different era classes. To this end, an learnable projection head is defined as $g_{\thetav}: \{\hv_a\} \rightarrow  \zv$, where $\hv_a$ is the encoded embedding from mel-spectrogram $\xb$, and $\zv$ is the latent representation.

%
      
For training, we first organize the training songs by their era labels. 
Let $I$ be the training set, $\zv_{i}$ be the latent representation for song $i$,  and $P(i)$ be the set with the same era label as song $i$'s, while $N(i)$ be the set with different labels from song $i$'s. The learning process is illustrated in Fig.~\ref{fig:bef_sup} and Fig.~\ref{fig:aft_sup}. Then, the EC loss can be defined as follows:
\begin{align}\label{eq:sup_cl}
   \mathcal L_{EC}= -\sum_{i \in I}\sum_{j \in P(i)} \text{log}\frac{\sigma(\zv_i \cdot \zv_j / \tau)}{\sum_{k \in N(i)}\sigma(\zv_i, \zv_k / \tau)},
\end{align}
where $\sigma(\cdot)$ is the non-linear transform which adopts an exponential function, and $\tau$ is the temperature parameter.
Note that $\mathcal L_{EC}$ brings two beneficial properties~\cite{khosla2020supervised}: $(\RN{1})$ ability to align all available positive samples; $(\RN{2})$ ability to mine hard positive and negative pairs. These properties can lead to a more robust representation of learning in our supervised scenario.

\subsection{Multi-Modal Contrastive (MMC) Learning}
\label{sec:audioart-mmc}

        
        
\begin{figure}
\centering
\includegraphics[width=\columnwidth, trim={0.1cm 0.1cm 0.1cm 0.1cm}, clip]{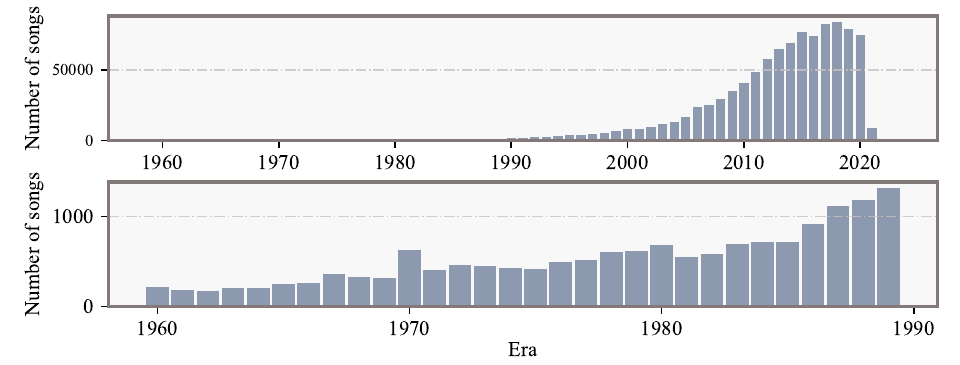}
\caption{Era distribution of our \tts{In-House} dataset. Upper sub-figure covers the years of from 1960 to 2020. Lower sub-figure enlarges the years of pre-1990.}
\label{fig:longtail}
\end{figure}

Fig. \ref{fig:longtail} presents the era distribution of an \tts{In-House} dataset across years, indicating a high imbalance between pre- and post-2000. In this ection, we propose the Multi-Modal Contrastive Learning to tackle this issue.

\subsubsection{Multi-Modal Fusion Module}

Multi-modal approaches have increasingly attracted researchers' attention owing to their promising results on various deep learning tasks \cite{ding2016audio, pini2017modeling}. To learn a better music representation, text about the music, such as lyrics, can be helpful augmented information and has been widely used in MIR tasks~\cite{lamere2008social,oramas2017multi,oramas2018multimodal}. However, given a song, its lyrics are not always available for reasons such as copyright issues. 
Alternatively, the text of the artist's biography contains rich information about the music style during the artist's active years~\cite{li2004music}. We choose the artist's biography considering its availability as opposed to lyrics.

\begin{figure}[t]
  \centering
  \subfigure[\label{fig:branch}\emph{Audio-Artist Encoder} ]{
      \includegraphics[height=5.1cm, trim={0.7cm 0.7cm 0.7cm 0.7cm}, clip]{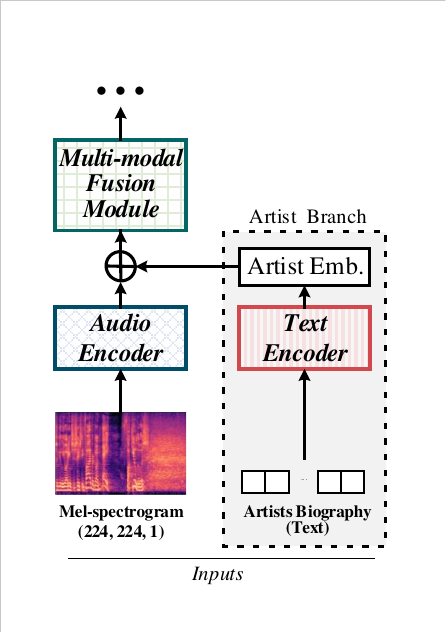}}
  \subfigure[\label{fig:contra}Contrastive Learning Framework]{
      \includegraphics[width=4.5cm, trim={0.7cm 0.5cm 0.7cm 0.7cm}, clip]{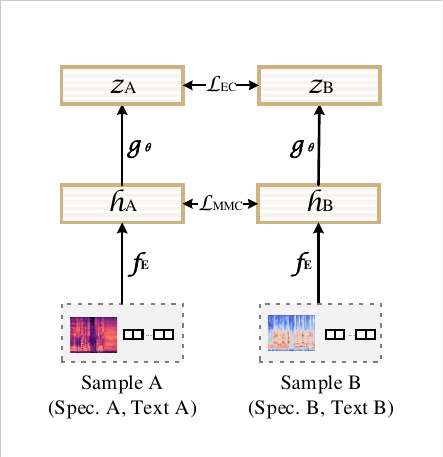}}\\
\caption{The overall illustration of \emph{AudioArt-MMC}. (a) is the \emph{Audio-Artist Encoder} mainly contains audio encoder, artist biography text encoder and a multi-modal fusion module. (b) shows the encoded latent representation $\zv$ that goes through corresponding projection heads for MMC loss and EC loss, respectively.}
\label{fig:backbone}
\end{figure}

We follow the previous work~\cite{simonetta2019multimodal} to make full use of contextual and correspondence information between music and text via attention mechanism, as illustrated in Fig. \ref{fig:backbone}(a). Given an input song $i$, we use two embedding vectors, $\ab_i \in \mathbb{R}^{d_h}$ and $\tb_i \in \mathbb{R}^{d_h}$, to represent its audio and artist biography information, respectively, and $d_h=64$ is adopted in the experiments. Specifically, $\ab_i$ is derived from the mel-spectrogram $\xb_i$ using the audio encoder. For $\tb_i$, we utilize Sentences-BERT \cite{reimers2019sentence} to encode the corresponding artist biography text. Then, we concatenate $\ab_i$ and $\tb_i$ to obtain $\sv_i \in \mathbb{R}^{2d_h}$ as the input of the self-attention layer.

The fusion module is based on Transformer blocks \cite{vaswani2017attention}. Specifically, the output of $l$-th Transformer block $\mathbf{ \Tilde{H}}_{\ell}$ is the concatenation of multiple attention heads: $\mathbf{\Tilde{H}}_{\ell} = [\mathbf{A}_{{\ell}, \text{1}}, \cdots, \mathbf{A}_{\ell, h}]$ where $h$ is the number of heads. Each head $\bf A$ contains trainable parameters $\mathbf{W^Q}_{\ell}, \mathbf{W}^K_{\ell},\mathbf{W}^V_{\ell} \in \mathbb{R}^{2d_h \times d_k}$, where $d_k$ is the projection dimension of the attention layer. 

\vspace{-.1cm}
\subsubsection{Multi-Modal Contrastive (MMC) Loss}
\label{subsec:cl}

The EC loss $\mathcal L_{EC}$ conditioning on the era labels directly helps improve the era classification task. However, it may excessively blur the differences of songs in the multi-modal embedding space within the same era class. Therefore, to avoid the representation collapse, we develop the multi-modal contrastive (MMC) loss, an unsupervised contrastive learning objective. The critical part for the effectiveness of MMC learning is a reasonable design of different \textit{views} of the input. For example, in computer vision, data augmentation techniques such as cropping and rotation are applied to one anchor image and the other to create positive and negative views of the anchor image. Similarly, we propose \textit{text-shuffle}, a technique to create multiple views for the combination of audio and artist embeddings input. To be specific, we consider the concatenation of matched audio and artist embeddings $\sv_{i,i} = [\ab_i, \tb_i]$ as a reference, so the concatenation of mismatched audio and artist creates negative views, e.g., $\sv_{i,k} = [\ab_i, \tb_{k}]$, where $\tb_{k}$ is a randomly sampled biography embedding from a different artist. 
To apply the MMC loss, two components are designed as follows:
\begin{itemize}
    \item \emph{Audio-Artist Encoder}: $f_{E}: \{\sv\} \rightarrow  \hv_m$, where $\hv_m$ is the multi-modal embedding learned from the audio-artist concatenation embedding $\sv$.
    \item \emph{Projection Head}: $f_{T}: \{\tb\} \rightarrow  \hv_m$, which projects the artist biography $\tb$ into the multi-modal embedding space.
\end{itemize} 
For simplicity, we take the artist biography as the anchor. Let an artist embedding $\tb_i$ be the anchor, its positive and negative examples in the audio-artist concatenation embedding space are $\sv_{i,i}$ and $\sv_{i,k}$, respectively.
The MMC loss is then defined as:
\begin{align}\label{eq:sup_cl1}
   \mathcal L_{MMC}= -\sum_{c \in \mathrm{C}}\sum_{i \in c(i)} \text{log}\frac{\sigma\left(f_{T}(\tb_i) \cdot f_{E}(\sv_{i,i}) / \tau\right)}{\sum_{k \in c(i)}\sigma\left(f_{T}(\tb_i) \cdot f_{E}(\sv_{i,k}) / \tau \right)},
\end{align}
where $\sigma(\cdot)$ is the exponential function, $\mathrm{C}$ is the all possible era labels, $c$ is the era label of song $i$, and $c(i)$ is the set having the era label $c$. The learning process is illustrated in Fig.~\ref{fig:bef_un} and Fig.~\ref{fig:aft_un}. As illustrated in Fig. \ref{fig:backbone}(b), we obtain the \emph{AudioArt-MMC} model, which is trained with a sum of the cross-entropy (MLE) loss, era contrastive (EC) loss, and multi-modal contrastive (MMC) loss:  
\begin{align}
    \mathcal L = \mathcal L_{\mathrm{MLE}} + \alpha \mathcal L_{MMC} + \beta \mathcal L_{EC}, \label{eq:loss}
\end{align}
where $\alpha$ and $\beta$ are hyperparameters weighting the loss terms. 
Note that, \emph{AudioArt-MMC} uses the multi-modal embedding for $g_{\theta}$ to calculate $\mathcal L_{EC}$ (see Eq. \ref{eq:sup_cl}), while \emph{Audio-SUC} uses the audio embedding encoded from mel-spectrogram for $\mathcal L_{EC}$.

\begin{figure*}[h]
\centering
\includegraphics[width=.8\textwidth, trim={.5cm .1cm 0.05cm .5cm}, clip]{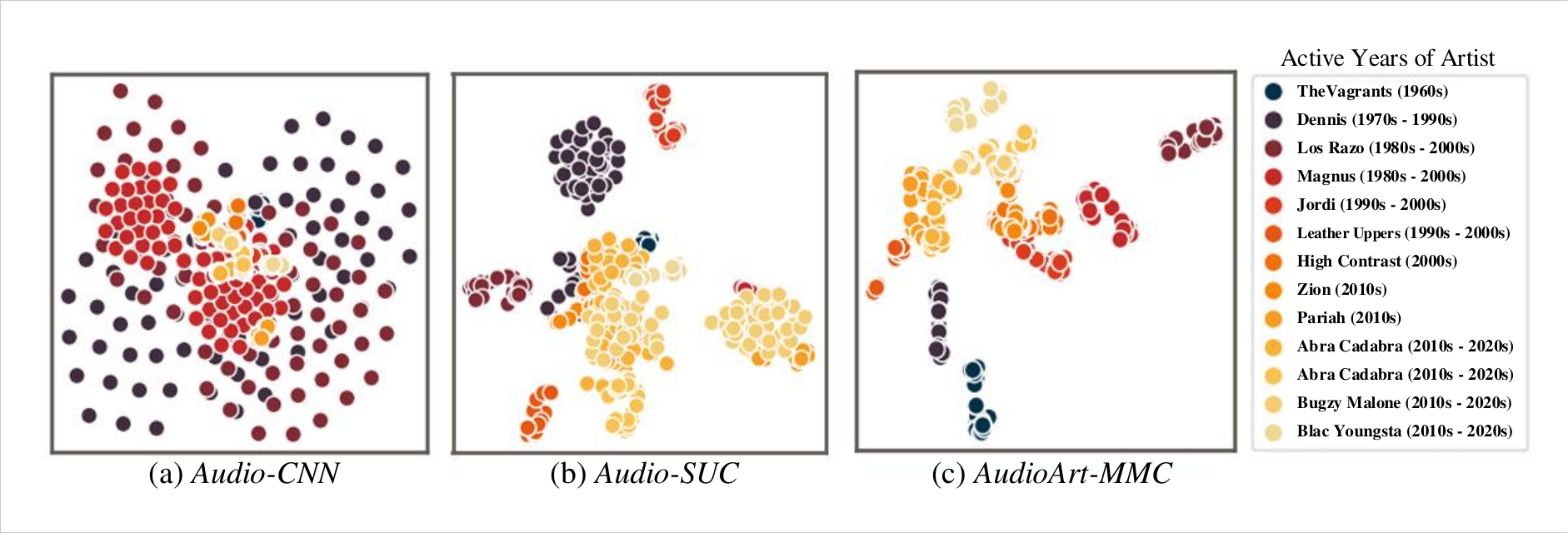}
\vspace{-.1cm}
\caption{The t-SNE visualization of the latent representation of different methods. The active years of an artist is marked in the parenthesis.}
\label{fig:tsne}
\end{figure*} 

\section{Experiments}
\label{sec:exp}

\subsection{Experiment Setup}

\textbf{Datasets \& Metrics.} To evaluate our proposed methods, we used a public dataset, Million Song Dataset (\tts{MSD}) \cite{bertin2011million}, and an internal music collection (\tts{In-House}).
MSD, which has been widely used in music auto-tagging evaluations \cite{nam2018deep}, contains the metadata and pre-computed audio features of one million contemporary songs. Our \tts{In-House} dataset includes about 800K songs, where the era distribution is presented in Fig. \ref{fig:longtail}. For artist biography, we adopted the AllMusic dataset \cite{malheiro2016emotionally}, which covers the biography text data of about 200K artists. We created the audio-text input pairs by matching the artist names. 
We selected 85,475 tracks from \tts{MSD} for the experiments, 
considering the availability of audio features, release years, and artist biographies. 
Whereas, the full set of \tts{In-House} was used, since it covers all the required information.

The difference in instrumentation and genre of songs released in close years
can be subtle. Therefore, we follow similar works~\cite{angulu2018age, saraf2022distribution} to propose a metric that allows false prediction tolerance. We define the accuracy with $\pm x$ years of tolerance (ACC$_x$) as follows: 
\begin{equation}
    \text{ACC}_x = \frac{1}{N} \sum_{i=1}^N \frac{x-\text{min}[\text{abs}(y_i-\hat{y_i}), x]}{x},
\end{equation}
where $y_i$ and $\hat{y_i}$ represent the ground truth and predicted years of song $i$, and $N$ is the total number of test songs. For instance, ACC$_0$ counts only the cases when the prediction exactly matches the ground truth. We define two evaluation scenarios of granularity for year ranges: $(\RN{1})$ \emph{one year per class}: each year represents a class. For example, songs released in 1981 are assigned with a different class than songs released in 1982. $(\RN{2})$ \textit{ten years per class}: each class represents a decade (e.g., 60s, 80s, and 00s). Therefore, songs released in 1981 and 1982 are within the `1980s' class, while the song in 2011 is labeled with `2010s' class.

\vspace{0.3cm}
\noindent \textbf{Training Details.} To extract the input mel-spectrograms, we first resample the original audio waveform to 22,050 Hz. Then, we use a window size of 2,048 with a hop size of 512 for a frame, and transform each frame into a 224-band mel-scaled magnitude spectrum. In the training stage, we set a batch size of 64. Each audio example, which includes 1,024 frames (roughly 6-seconds long), is randomly sampled from an arbitrary song in the training data. We use the Adam optimizer \cite{kingma2014adam} with a learning rate of 1e-4.

\begin{table}[t]
\centering
\tabcolsep=1pt
\renewcommand\arraystretch{1.1}
\begin{tabular}{l|cccc|cccc}
   \toprule[1pt]
    &\multicolumn{4}{c|}{\texttt{MSD}}&\multicolumn{4}{c}{\texttt{In-House}}\\
    \textbf{Method}&\footnotesize{ACC}$_{0}$&\footnotesize{ACC}$_{1}$&\footnotesize{ACC}$_{2}$&\footnotesize{ACC}$_{3}$&\footnotesize{ACC}$_{0}$&\footnotesize{ACC}$_{1}$&\footnotesize{ACC}$_{2}$&\footnotesize{ACC}$_{3}$\\\hline\hline
    \multicolumn{9}{c}{64 classes (\textit{one year per class})}\\\hline
    \emph{AudioArt-MMC}             &\textbf{.238}&\textbf{.451}&\textbf{.588}&\textbf{.628}&\textbf{.188}&\textbf{.350}&\textbf{.476}&\textbf{.650}\\
    \emph{Audio-SUC}     &.213&.313&.426&.544&.140&.196&.436&.579\\
    \emph{Audio-CNN}    &.138&.169&.271&.302&.097&.157&.334&.486\\\hline
    \multicolumn{9}{c}{8 classes (\textit{ten years per class)}}\\\hline
    \emph{AudioArt-MMC}             &\textbf{.788}&\textbf{.975}&\textbf{.988}&-&\textbf{.750}&\textbf{.978}&\textbf{.986}&-\\
    \emph{Audio-SUC}     &.675&.917&.958&-&.563&.920&.973&-\\
    \emph{Audio-CNN}    &.519&.902&.964&-&.433&.865&.954&-\\
    \bottomrule
\end{tabular}
\caption{Results of the \textit{one year/class} and \textit{ten year/class} are reported. Best results of each dataset and metric are marked \textbf{bold}.}
\vspace{-0.3cm}
\label{tab:comp}
\end{table}    

\subsection{Results and Discussion}

We performed 10-fold cross-validation to obtain the results, and this process was done on \tts{MSD} and \tts{In-House} independently. We split 10\% of the training set as the validation set, and selected the snapshot with best performance on validation set as the testing model.
    


\vspace{0.3cm}
\noindent \textbf{Era Prediction Comparison.} Table~\ref{tab:comp} presents the overall comparison of the two datasets. It is clear that incorporating the supervised contrastive learning (\emph{Audio-SUC}) can significantly outperform the baseline model (\emph{Audio-CNN}) for most of the cases (except the ACC$_2$ in the 8 classes scenario on \tts{MSD}), and that jointing the audio and artist information for input with the multi-modal contrastive learning (i.e., \emph{AudioArt-MMC}) can consistently further the performance. Such a result is in line with our expectations when designing the models. As suggested in prior work \cite{interiano2018musical}, artist information is very helpful in hit song prediction of a period of time, so it also offers strong clues about the musical sounds representing an era. 
Comparing among ACC values with different tolerances for the 64 classes scenario, the superiority of the contrastive learning-based models over CNN seems to become more apparent when the tolerance is getting larger. This indicates that the learned features indeed possess abilities for era discrimination. Lastly, for the 8 classes scenario (i.e., decade tagging), ACC$_1$ can achieve 90\%+ accuracy solely based on audio, demonstrating promising results for a usable system in relevant MIR applications. 

\vspace{0.3cm}
\noindent \textbf{Robustness to Data Imbalance}. As discussed, our dataset is imbalanced in era distribution. We are interested in the performance of the pre-2000 era, so we present the result of the corresponding subset in \tts{In-House}. 
Table~\ref{tab:longtail} shows the results on the pre-2000 subset and the difference $\Delta$ from their counterparts in Table~\ref{tab:comp}. Note that smaller $\Delta$ is better. Comparing the $\Delta$ values among the three variants, \emph{AudioArt-MMC} shows significantly smaller performance gaps between the full-set and subset results, indicating its stronger ability to handle the imbalanced data. 


\vspace{0.3cm}

\noindent \textbf{Visualization of Embeddings.} Finally, we perform a qualitative examination of the effect of the proposed MMC loss by visualizing the embeddings of the penultimate layer from the three proposed models using t-SNE plots. We randomly select 400 songs associated with 13 artists and plot the corresponding song embeddings in the 2D space. Each artist is assigned a unique color, with brighter colors for recenter eras. As displayed in Fig.~\ref{fig:tsne}, EC loss can facilitate the embeddings of same era class to be closer while maintaining the diversity of different songs (see Fig.~\ref{fig:tsne}(b)), and MMC loss further forces aggregating songs of the same artist (see Fig.~\ref{fig:tsne}(c)). 

\begin{table}
\centering
\tabcolsep=1.2pt
\begin{tabular}{l|cccc|cccc}
\toprule[1pt]
\textbf{Method}&\footnotesize{ACC}$_{0}$&\footnotesize{ACC}$_{1}$&\footnotesize{ACC}$_{2}$&\footnotesize{ACC}$_{3}$& $\Delta_{0}$& $\Delta_{1}$& $\Delta_{2}$& $\Delta_{3}$\\\hline\hline
\emph{AudioArt-MMC}             &\textbf{.175}&\textbf{.287}&\textbf{.393}&\textbf{.488}&\textbf{.013}&\textbf{.063}&\textbf{.083}&\textbf{.162}\\\hline
\emph{Audio-SUC}     &.094&.122&.237&.379&.046&.074&.097&.200\\
\emph{Audio-CNN}     &.055&.093&.162&.295&.042&.064&.172&.191\\
\bottomrule
\end{tabular}
\caption{Results for the pre-2000 subset of \tts{In-House} in the 64 classes (one year/class) scenario. $\Delta_i$ corresponds to the ACC$_i$ difference between the pre-2000 subset and the whole dataset.}
\label{tab:longtail}
\vspace{-0.3cm}
\end{table}    

\section{Conclusion}

In this paper, we have presented a novel multi-modal contrastive learning framework to recognize music era based on audio and artist information. The design of the EC and MMC losses enables a satisfactory clustering behaviors, showing convincing results for the task. 
For future work, we will try to improve the performance with more sophisticated solutions. We also plan to study the possibility of including additional metadata, such as instrumentation, genre, and mood, into a unified framework.

\section{Acknowledgement}
    \vspace{-.05cm}	
    This work was supported by National Key R\&D Program of China (2019YFC1711800) and NSFC (62171138).

\bibliographystyle{IEEEbib-abbrev}
\bibliography{refs}
\end{document}